# Computational Identification and Stuart-Landau Modeling of Collective Dynamical Behaviors of Octuple Laminar Diffusion Flame Oscillators

Tao Yang[1], Yuan Ma[1], Peng Zhang[2,*]

*1. Department of Mechanical Engineering, The Hong Kong Polytechnic University, Hung Hom, Kowloon, Hong Kong*

*2. Department of Mechanical Engineering, City University of Hong Kong, Kowloon Tong, Kowloon, Hong Kong*

## ABSTRACT

Annular chambers, consisting of multiple flame nozzles, are frequently used in many industrial processes, for example, rocket engines and gas turbines. In the study, we proposed a novel approach to the problem of annular combustion with emphasis on the collective dynamical behaviors that its individuals do not have. A series of circular arrays of octuple flickering laminar buoyant diffusion flames were investigated computationally and theoretically. Five distinct dynamical modes, such as the merged, in-phase mode, rotation, flickering death, partially flickering death, and anti-phase modes, were computationally identified and interpreted from the perspective of vortex dynamics. A unified regime diagram was obtained in terms of the normalized flame frequency $f/f_0$ and the combined parameter $(\alpha - 1)Gr^{1/2}$, where $\alpha = l/D$ is the ratio of the flame separation distance $l$ to the flame nozzle diameter $D$ and $Gr$ is the Grashof number. The bifurcation transition from the in-phase mode and the anti-phase mode to the totally or partially flickering death occurs at $(\alpha - 1)Gr^{1/2} = 655 \pm 55$. In addition, a Stuart-Landau model with a time-delay coupling was utilized to reproduce the general features and collective modes of the octuple oscillators flame systems.

## KEYWORDS

---

[*] Corresponding author
 E-mail address: penzhang@cityu.edu.hk, Tel: (852)34429561.

# I. INTRODUCTION

Collective behaviors of a dynamical system refer to the patterns and phenomena that emerge when individual components or agents interact and synchronize [1, 2]. These dynamical behaviors are studied across various scientific disciplines, including physics, biology, chemistry, sociology, and engineering [3]. A type of thermal fluid system with rotational symmetry [4-8], is common in many industrial devices, for example, turbomachine rotors [9], can-annular combustors in gas-turbine engines [10, 11], and Chevron-type primary exhaust nozzles for aircraft engines [12]. A prominent example that inspires the present study is the annular combustion system, where multiple combustion chambers are arranged in a circle array, and the collective dynamics of these multi-element flames is attributed to the potentially destructive azimuthal instability [?]. A simplified but potentially vaulable study of the problem is to investigate the dynamical modes of circulary arrays of laminar diffusion flame oscillators, which retain the essential features of the system but theoretrically and computationally much tractable.

A well-known self-exciting flame oscillator is the "flickering" or "puffing" buoyant diffusion flame[13]. Many previous studies [14-23] have been carried out to understand the physics of flickering diffusion flames. Chen et al. [24] confirmed that the large toroidal vortices (e.g., vortex rings) are formed outside the luminous flame due to the buoyance-induced instability, as shown in Fig. 1(a), and found that the frequency of the toroidal vortices well correlates with the flicker frequency. In early experimental studies, the flicker frequency $f_0$ is proportional to $(g/D)^{1/2}$, where $g$ is the gravitational constant and $D$ is the fuel inlet diameter. The flickering phenomenon was observed and studied in diverse research areas of fluid mechanics (i.e., the buoyant jets and plumes [25, 26]), flame dynamics (i.e., the wake flame of droplets and porous spheres [27, 28], nonnormal environments of gravity and pressure [15, 18], and co-flow and swirling effects [29, 30]), chemical reaction (i.e., soot formation [31]), and flames with acoustic, electric, and magnetic forcing [32-34]. Therefore, flickering flames is a simple and fundamental platform for practical issues and multidisciplinary research.

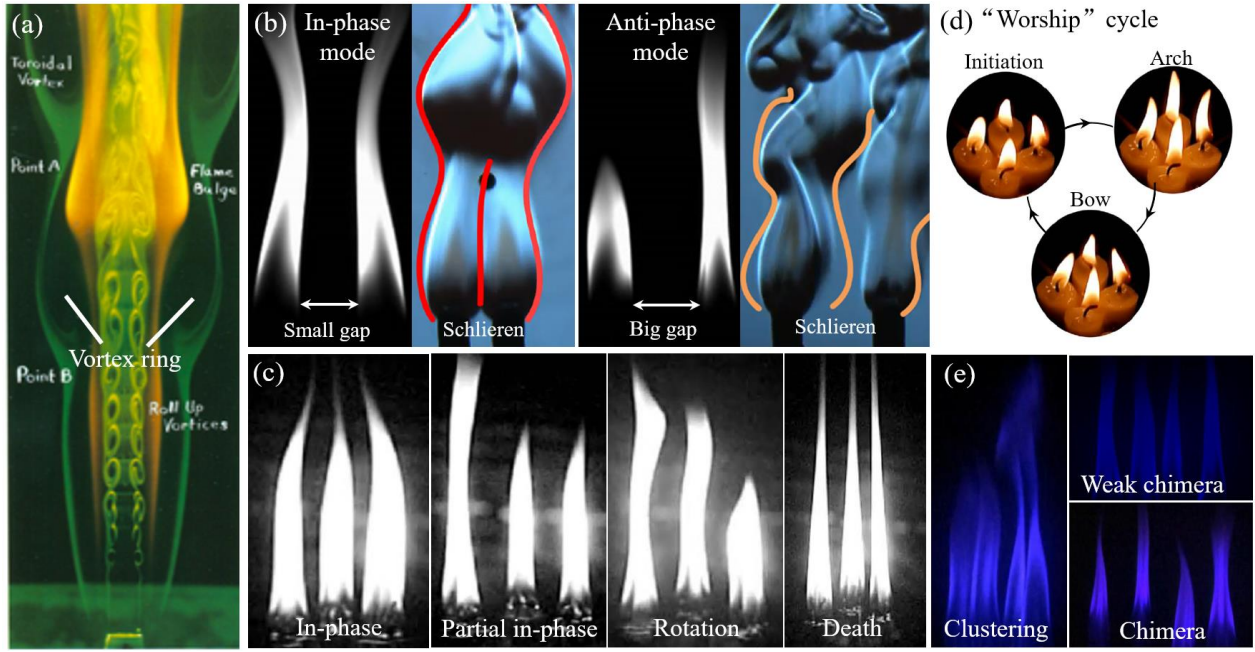

FIG. 1. (a) Jet diffusion flame with the outer vortex ring (e.g., toroidal vortex) [24], (b) two dynamical modes of dual flickering flame [35], (c) four dynamical modes of triple flickering flames [36], (d) initial-arch-bow-initial "worship" oscillation mode of four candle flames [37], and (e) three dynamical states, namely clustering, weak chimera, and chimera, of four flame oscillators [38].

More is different [39]. The individual flickering flames typically follow simple rules of a self-exciting oscillation, but their collective interactions can lead to complex and emergent properties. In the past decade, two dynamical modes of dual flickering flames were substantiated by numerical simulations [40-42] and experiments [43, 44]. The synchronized flames flicker in an in-phase manner at a relatively small gap while they do in an anti-phase manner at a relatively large gap, as shown in Fig. 1(b). Recently, the dynamical behaviors of multiple flickering flames (i.e., more than two identical flame oscillators) have attracted much research interest [36-38, 45, 46]. Okamoto et al. [36] observed four distinct dynamical modes, such as the in-phase mode, the partial in-phase mode, the rotation mode, and the death mode, in three flickering candle flames in an equilateral triangle arrangement, as shown in Fig. 1(c). Chi et al. [47] systematically studied triple flickering flames in an isosceles triangle arrangement and recognized seven distinct stable dynamical modes. Forrester [37] experimentally observed an initial-arch-bow-initial "worship" oscillation mode for four candles in Fig. 1(d). Manoj et al. [38, 45] experimentally observed variants of clustering and chimera states, for example, three dynamical states, namely clustering, weak chimera, and chimera, of four candle-flame oscillators in Fig. 1(e). Therefore, interacting flickering flames can give rise to rich dynamical phenomena. Moreover, the flame number and arrangement are essential to the complexity of flame dynamics.

Compared with the many experimental studies on multiple flickering buoyant diffusion flames, very few studies attempted to establish dynamical models to reproduce the experimental findings. By hypothesizing that a lack of oxygen is a key factor in producing the flickering flame and that thermal radiation coupling causes the synchronization of two flames, Kitahata et al. [43] proposed a dynamical model to interpret their experiments. Noting that their radiation measurement does not support the radiation coupling hypothesis, Gergely et al. [48] hypothesized that the oxygen flow induced by the thermal expansion is responsible for the flame coupling and proposed a modified model. Manoj et al. [49] adopted the time-delay coupled identical Stuart-Landau oscillators to reproduce the dynamical modes of two coupled candle flames. Recently, Chi et al. [50] proposed a complex coupling term in the Stuart-Landau equation to interpret the experimentally identified dynamical modes. They also found that the classical Kuramoto model successfully interpreted the dynamical modes except for those associated with amplitude death while the complexified Stuart-Landau model well interpreted all the dynamical modes observed in their experiment.

Regardless of the above advances, the existing studies are often limited to a relatively small dynamical system consisting of a small number of flames. Few researchers have yet carried out numerical simulations and theoretical modeling on larger flame systems. To make a small but firm step toward understanding the complex combustion dynamics of annular combustion systems, this study attempts to computationally and theoretically study the collective dynamical behaviors of octuple laminar diffusion flame oscillators. We shall organize the remaining text along the lines described here. The simplified laboratory problem for investigating combustion modes of annular combustors is first presented in Sec. II. Then, the methodology descriptions of computation for circular arrays of multiple flickering flames and the dynamical oscillator model for flame flicker are given in Sec. III. Next, five dynamical modes of circular systems of octuple flames are provided and interpreted from the vortex dynamical perspective, followed by the discussion on the dynamical mode transition, in Sec. VI. The last section contains a summary of the present findings, significance, and potential applications in Sec.V.

## II. PROBLEM FORMULATION: A NOVEL APPROACH TO COLLECTIVE DYNAMICS OF ANNULAR COMBUSTION

Juniper and Sujith [51] reviewed the modeling development for rocket engines and gas turbines and pointed out that physical understanding and parametric learning from laboratory-scale rigs are beneficial. In recent years, combustion instability manifested as azimuthal mode coupling in the annular chambers of many gas turbines has received increasing attention [52-56]. As shown in Fig. 2(a), the specific mode propagates in the azimuthal direction and not in the longitudinal direction [57]. The azimuthal instability originates from the collective behavior of multiple flames and is not suitable for the simplification study in that the integrated system is separated by a single or a few flame modules with appropriate periodic boundary conditions. While turbulent flames are the most common combustion phenomena in nature and industries, they have complex flow structures and strong turbulent/chemical couplings. Resolving multiscale spatiotemporal features of turbulent flames requires significant computational resources. Vignat et al. [6] found that laminar flames, created by the matrix injectors in Fig. 2(b), are useful for investigating the annular systems coupled by azimuthal modes in the absence of swirl and turbulent fluctuations.

The extreme conditions and the complex physics in combustors make it difficult to perform full-scale experiments, or at least to carry out detailed measurements [57]. Therefore, simple tools are usually required to reproduce instability modes, comprehend their physics, and evaluate control strategies. Previous works [52, 53, 58-60] have attempted to construct analytical models for the prediction and control of combustion instabilities while retaining most of the important physical phenomena and geometrical specificities of annular chambers. Parmentier et al. [52] established a simple analytical model based on a network view of the annular chamber to predict stability maps for the azimuthal modes. Bonciolini et al. [59] proposed low-order models consisting of coupled self-sustained oscillators driven by additive stochastic forcing to qualitatively reproduce the experimentally observed dynamics. Weng et al. [60] presented a phenomenological, reduced-order model based on a synchronization framework for modeling the transition between different dynamical states. It is noteworthy that classifying various instability modes in annular chambers is an interesting and valuable issue [53].

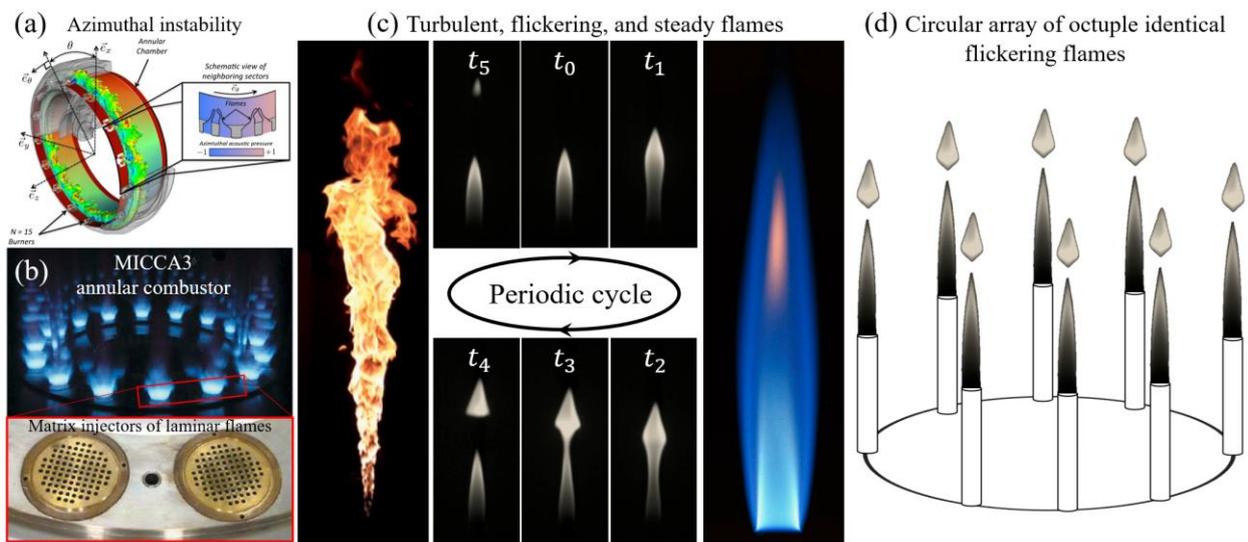

FIG. 2. (a) Azimuthal combustion instability (pressure fluctuations along the azimuthal direction) in an annular engine and zoom on two neighboring burners [57], (b) MICCA3 annular combustor with matric injectors [6], (c) turbulent [61], flickering [47], and steady [62] flames, and (d) circular array of octuple flickering flame.

The present strategy for studying the dynamical modes in the annular chambers is different from the above ones. It is well known that turbulent flames are ubiquitous but difficult to study due to complex turbulent flow and strong turbulent and chemical coupling, while laminar flames are more tractable but barely useful in industries due to too simple configuration and limited couplings of flow and reaction. To facilitate understanding of the collective dynamics of multiple flame systems in annular chambers, this study is mainly focused on unconfined laminar diffusion flames to minimize the complexity of turbulent fluctuations, combustion chemistry, and flame-acoustics interaction. Compared with complex turbulent flames and simple steady flames in Fig. 2(c), flickering flames [13] with unstable, nonlinear, and periodic characteristics retain the intrinsically unsteady nature of turbulent flames most simply. Moreover, the circular array of flickering flames, as shown in Fig. 2(d), could be a tractable research platform to grasp the essential features of flame-vortex and vortex-vortex interactions in turbulent combustion, as the circular flame systems possess annular flame configuration, nonlinear flame-flame interaction, and coupling of flame and vortex.

## III. COMPUTATIONAL AND THEORETICAL METHODOLOGY

### A. Computational Methods

Present simulations are performed using Fire Dynamics Simulator (FDS) [63], which is a finite-difference open-source solver developed by the National Institute of Standards and Technology (NIST). The solver platform is based on the "low Mach number" combustion equations [64] for describing the low-speed motion of a gas driven by chemical heat release and buoyancy forces [65, 66]. In the past decade, FDS has been widely used for studying thermally-driven flow issues, including the soot model of laminar flames [67], small-scale fire whirls [68-70], large eddy simulation of turbulent buoyant flames [71-73], and nonlinear dynamics of turbulent flames [41, 74, 75], to name only a few. Our recent works on single flame in rotating flow [76], dual interacting flames [42], and triple interacting flames [77] have proven the reliability of this computational platform in successfully reproducing flame and vortex interactions of laminar buoyant diffusion flames for distinct dynamical behaviors in experimental observation [36, 43, 78-80].

In the present study, the computational methods are similar to the previous studies of dual-flame [42] and triple-flame [77] systems. The laminar non-premixed methane-air flames are simulated by solving the governing equations of continuity, momentum, and energy conservations:

$$\frac{\partial \rho}{\partial t} + \nabla \cdot (\rho \boldsymbol{u}) = 0 \tag{1}$$

$$\frac{\partial}{\partial t}(\rho \boldsymbol{u}) + \nabla \cdot (\rho \boldsymbol{u}\boldsymbol{u}) = -\nabla \tilde{p} - \nabla \cdot \boldsymbol{\tau} + (\rho - \rho_\infty)\boldsymbol{g} \tag{2}$$

$$\frac{\partial}{\partial t}(\rho h_s) + \nabla \cdot (\rho h_s \boldsymbol{u}) = \frac{D\bar{p}}{Dt} + \dot{q}''' - \nabla \cdot \dot{\boldsymbol{q}}'' \tag{3}$$

where $\rho$ is the mass density and $\boldsymbol{u} = (u, v, w)$ the velocity vector; the spatially and temporally resolved pressure consists of the pressure perturbation $\tilde{p}$ and the backpressure $\bar{p}$, $\boldsymbol{\tau}$ the viscous stress tensor, $\rho_\infty$ the background air density, and $\boldsymbol{g} = (0,0,-g)$ the gravitational acceleration vector; $h_s$ is the sensible enthalpy (a mass-weighted average of the enthalpies of the individual gas species), $\dot{q}'''$ the combustion heat release rate per unit volume, and $\dot{\boldsymbol{q}}''$ the conductive and diffusive heat fluxes. Particularly, Eq. (3) is not solved explicitly but satisfied through guaranteeing the velocity divergence, which is factored out as follows:

$$\nabla \cdot \boldsymbol{u} = \frac{1}{\rho h_s}\left[\frac{D}{Dt}(\bar{p} - \rho h_s) + \dot{q}''' - \nabla \cdot \dot{\boldsymbol{q}}''\right] \tag{4}$$

where $h_s$ is calculated by summing $h_{s,i} = TY_i c_{p,i}$ of all gas species, where $T$ is the temperature, $Y_i$ the mass fraction of specie $i$, and $c_{p,i}$ the specific heat of specie $i$ at constant pressure; $\dot{q}'''$ is calculated by summing $\dot{m}_i''' \Delta h_{f,i}$, where $\dot{m}_i'''$ is the mass production rates of specie $i$ and $\Delta h_{f,i}$ is the respective heats of formation; $\dot{\boldsymbol{q}}''$ contains $-k\nabla T$ and $-\sum_i \rho h_{s,i} D_i \nabla Y_i$, where $k$ the thermal conductivity and $D_i$ is the diffusivity of species $i$. In addition, the background pressure is retained in the equation of state $\bar{p} = \rho \mathcal{R} T / \overline{W}$ (ideal gas law), where $\overline{W}$ is the molecular weight of the gas mixture, $\mathcal{R}$ the universal gas constant, and $T$ the temperature. Finally, the transport equations of specie $Y_i$ need to be solved for closure of the governing equations.

For the thermally driven flow of flickering flame, the buoyancy is predominant and the dynamic structures of flames are unaffected by the fast-chemistry assumptions [81]. Our recent works [30, 76] compared infinitely fast chemistry and one-step finite rate chemistry for reproducing flickering flames and confirmed that different chemistry models have slight influences on the flickering frequency. Thus, to avoid the complexity and high computation cost of a detailed reaction mechanism, the mixing-limited, infinitely fast reaction was used to calculate species mass fractions and the heat release rate. Specifically, the single-step reaction of methane and air is adopted to simplify the combustion in buoyancy-dominated flames. The transport equations of five gas species

($CH_4$, $O_2$, $CO_2$, $H_2O$, and $N_2$) are solved explicitly by

$$\frac{\partial}{\partial t}(\rho Y_i) + \nabla \cdot (\rho Y_i \boldsymbol{u}) = \nabla \cdot (\rho \mathcal{D}_i \nabla Y_i) + \dot{m}_i''' \tag{5}$$

where $Y_i$ is the mass fraction of specie $i$, $\mathcal{D}_i$ the diffusion coefficient, and $\dot{m}_i'''$ the mass production rate per unit volume by chemical reactions. To ensure the realizability of species mass fractions (i.e., $Y_i \geq 0$ and $\sum Y_i = 1$), FDS's strategy is to solve all species equations for $\rho = \sum(\rho Y)_i$ and then to obtain mass fraction by $Y_i = (\rho Y)_i/\rho$. Specifically, the primitive species are lumped into reacting groups, such as $Fuel + Air \rightarrow Products$ and the lumped species approach (a simplified reaction progress variable approach [82]) is used to avoid any complications related to boundedness and ill-defined initial and boundary conditions. See Supplemental Material [83], Sec. S1, for more details on the simplified approach. Besides, the effects of turbulence, radiation, and soot were not considered in the present small-scale laminar flames [76].

The basic solution procedure of the governing equations is a predictor-corrector explicit-time integration scheme for capturing unsteady and dynamic processes in thermally driven buoyant flows. All the spatial derivatives of the governing equations are discretized by the finite difference method on structured, uniform, staggered grids. The time step is adjusted to ensure numerical stability by checking the Courant–Friedrichs–Lewy (CFL) condition at the end of the prediction step. The solver algorithm has evolved over roughly three decades. Further details of the solver and a wide array of validation/verification applications can be found in [63].

### B. Computational Setups

In this study, the circular systems of octuple flames are studied by simulating eight identical Bunsen-type diffusion flames in circle arrays in a quiescent open space, as shown in Fig. 3. The square column of $16D \times 16D \times 24D$ for the present computational domain, where $D = 10$ mm is the characteristic length of the Bunsen burner. The Bunsen burners (the gray square columns of $D \times D \times 3D$) are located with the same gap distance of $l$ along a cycle of $L = \beta D$ diameter and labeled by 1-8 numbers in a clockwise direction. The bottom of each burner is an inlet boundary for the gaseous methane jet (density $\rho_F = 0.66$ kg/m³ and kinematic viscosity $\nu_F = 1.65 \times 10^{-5}$ m²/s) at the uniform velocity $U_0$, while the burner wall (grey area) is set as an impermeable, non-slip, and adiabatic solid boundary. The six sides of the computation domain are set as an open boundary condition (i.e., a simple upwind boundary condition). On these boundaries, the local pressure gradient determines whether gases flow inwards or outwards [76]. When the flow is incoming or outgoing, the temperature and species mass fractions take on their respective exterior values or the respective values in the grid cell adjacent to the boundary. The environment is set as quiescent air (density $\rho_A = 1.20$ kg/m³ and kinematic viscosity $\nu_A = 1.51 \times 10^{-5}$ m²/s) under the normal temperature and pressure conditions (20°C and 1 $atm$). The flames of all cases were in a fully developed state as the simulation time was at least 20 times longer than the characteristic time $L/U_0$. Table I shows the detailed setups for simulating the flickering phenomenon.

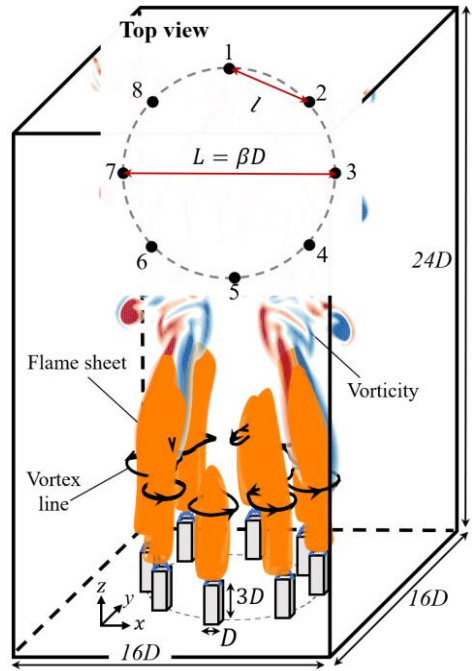

FIG. 3. Schematic of the three-dimensional simulation and flame arrangement (top view) for circular arrays of octuple identical flickering flames, labeled by 1-8 numbers in a clockwise direction. Each Bunsen burner (denoted by the grey squares) is consisted of a square column of $D \times D \times 3D$ and the centers of eight burners are located along a $L$ diameter circle with same distance of $l$. The whole computational domain is a square column of $16D \times 16D \times 24D$ with an open space boundary on each side. The octuple identical flickering flames (denoted by the orange flame sheets) are ejected from the burners, their respective toroidal vortices (denoted by the black vortex lines and the vorticity contours) around the flames interact each other, and the octuple-flame system presents distinct collective dynamics.

TABLE I. Key parameters of numerical simulations.

| Setup | Detailed description |
|---|---|
| Fuel and oxidizer | Methane gas with kinematic viscosity $\nu_F = 1.65 \times 10^{-5}$ m²/s is ejected at a uniform inlet velocity $U = 0.165$ m/s from each burner; Environment air with kinematic viscosity $\nu_A = 1.51 \times 10^{-5}$ m²/s is quiescent under the normal temperature and pressure conditions (20°C and 1 $atm$). |
| Bunsen burner | Identical square burners with $D = 10$ mm length and $3D$ height. |
| Nondimensional parameter | Reynolds number defined by $Re = UD/\nu_F = 100$ and Froude number defined by $Fr = U^2/gD = 0.28$. |
| Domain | A square column of $16D \times 16D \times 24D$ with the gravity constant of $g$=9.8 m/s². |
| Boundary | Open boundary conditions allow airflow in and out freely for six surfaces of the domain; impermeable, non-slip, and adiabatic boundary conditions for solid surfaces of the burners with the central fuel inlet. |
| Grid | Structured, uniform, and staggered grids of $\Delta x = \Delta y = \Delta z = 10^{-1}D$. |
| Model | Infinitely fast reaction for one-step overall methane/air combustion; no modeling for radiation and soot formation. |
| Simulation | Fully developed flows after running at least 20 times longer than $(L/U_0)$. |

It should be noted that the flickering feature of the flames results from the periodic shedding of toroidal vortices, which is induced by the buoyancy force. Many previous experiments have substantiated that the dominant flickering frequency is insensitive to the type of fuel [16, 22, 24, 62]. Therefore, a complex combustion mechanism is not indispensable for computationally reproducing the flame flicker. For simplicity of simulating the buoyancy-dominated diffusion flames, one-step overall methane/air combustion of $CH_4 + 2(O_2 + 3.76N_2) \rightarrow CO_2 + 2H_2O + 7.52N_2$ is adopted and the chemical reaction is mixing-limited, infinitely fast. Also, the radiation and soot formation were neglected in the small-scale flames.

### C. Domain and Grid Independent Study

Our previous mesh-independence study of flickering flames [42] shows that the mesh refinement (each grid has $\Delta x/D = \Delta y/D = \Delta z/D = 10^{-1}$) is sufficient to capture the primary vortical structure of the buoyance-induced flicker of a laminar diffusion flame. In the present study, we take the case of $L = 10.2D$ as an example and carry out a domain-independence study. The flame dynamics will be elaborated in the following sections. Table II compares the dynamics mode of flames and flickering frequency obtained in four cases with different domains and grids (see Supplemental Material [83],

Fig. S1, for more details). The mode is determined by the phase difference, while the dominant frequency is calculated by the Fast Fourier Transform of flame signals (e.g., the vertical velocity at a fixed point [76]) with an $\pm 0.2$ Hz error.

TABLE II. The domain and grid-independence studies.

| No. | Domain and grid | Flame mode | | Frequency (Hz) |
|---|---|---|---|---|
| 1 | $20D \times 20D \times 30D$ $200 \times 200 \times 300$ | Anti-phase: two groups with $\pi$ phase difference | 1-3-5-7 in-phase; 2-4-6-8 in-phase | 12.2 12.2 |
| 2 | $16D \times 16D \times 24D$ $160 \times 160 \times 240$ | Anti-phase: two groups with $\pi$ phase difference | 1-3-5-7 in-phase; 2-4-6-8 in-phase | 12.4 12.4 |
| 3 | $16D \times 16D \times 24D$ $240 \times 240 \times 360$ | Anti-phase: two groups with $\pi$ phase difference | 1-3-5-7 in-phase; 2-4-6-8 in-phase | 12.8 12.8 |
| 4 | $12D \times 12D \times 18D$ $120 \times 120 \times 180$ | Partially flickering death: two groups with different dynamics | 2-6 and 4-8 in-phase; 1-3-5-7 flickering death | 12.4 15.8 |

The present results show that the second domain and grid study can capture well the flame mode and frequency, as the same mode and a small frequency change (about 5%) are simulated by enlarging the domain and refining the mesh. Consequently, we adopted the domain of $16D \times 16D \times 24D$ and the grid of $160 \times 160 \times 240$ for the parametric studies in the following sections to ensure adequate accuracy with reasonable computational cost. According to the thermal diffusivity (about 20-140 mm$^2$/s at 300-2100 $K$) and the characteristic time of the present flickering flames (flame frequency about 10 Hz), the diffusion zone of the flame is evaluated to be 1.4-3.7 mm and can be partially resolved by the present mesh size [76].

### D. Validations of Single and Dual Flickering Flames

Based on vortex dynamics, Xia and Zhang [21] theoretically established the connection between the flicker of a single flame and the periodic shedding of toroidal vortices. Their vortex-dynamical scaling theory for flickering buoyant diffusion flames was very well validated by the existing experimental data in the literature. As a qualitative validation of the present computation, a benchmark case of a diffusion methane flame at $Re = 100$ is simulated to reproduce the flickering phenomenon, where flame dynamics is closely associated with vortex evolution. For more details on the evolution of the temperature and vorticity contours during a period, please see Sec. S3.1 and Fig. S2 in Supplemental Material [83]. Particularly, the flame mainly flickers in the vertical direction and its representative quantity (e.g., the temperature in the transverse section) is time-varying like a periodic wave. The periodic flicker of a diffusion flame essentially exhibits a limit cycle [23, 48, 49], which could be modeled as a Stuart-Landau oscillator. The relevant discussion will be expanded in Sec. III.E.

Many early experimental studies [14, 18, 84] empirically obtained the correlation of $St \sim Fr^{-1/2}$ for flame flicker, where $Fr = U^2/gD$ is the Froude number and $St = f_0 D/U$ is the Strouhal number. To quantitatively validate the present computational methodology and models, we compared the flickering frequency of single flames with previous experiments (see Supplemental Material [83], Sec. S3.2 and Fig. S3, for more validation details). It is shown that the numerical results agree fairly with the scaling relation of $St \sim Fr^{-1/2}$, while the experimental data scatter within a strip between $St = 0.28 Fr^{-1/2}$ and $0.54 Fr^{-1/2}$. Overall, the present simulations are quite capable of capturing the vortex evolution and flame flicker in single diffusion flames in a quiescent environment. In addition, we computationally and theoretically investigated the flame flicker in swirling flows and found the nice reliability and accuracy of the present simulation framework in reproducing previous experimental observations, including the faster flicker of diffusion flames in weakly swirling flows [76] and various dynamical modes of diffusion flame in the environment with a range of strong swirls [30]. For capturing the vortex interactions between flickering flames, the present computational method has been sufficiently validated in the following.

The interactions of two identical flames are fundamental to understanding the bifurcation of various dynamical modes of circular arrays of flame oscillators. Over a decade, the coupling mechanism of two flame oscillators has been extensively investigated in terms of flow dynamics [29, 42], flickering modes [35, 49], and reduced-order modeling [43, 48]. Considering the interaction of two toroidal vortices in a dual-flame system, Yang et al. [42] used the vortex diffusion and the vortex-induced flow to interpret the in-phase and anti-phase flickering modes occurring at relatively small and big flame distances, respectively. See Supplemental Material [83], Sec. S4 and Fig. S4, for more details on the vortex mechanism. Particularly, Yang et al. [42] found that the synchronized flickering mode of two identical flames is not dictated by the nondimensional gap distance $\alpha = l/D$, where $l$ is the distance between the centers of two identical flames and $D$ is the characteristic size of burner nozzle. Then, a characteristic nondimensional number $(\alpha - 1)Gr^{1/2}$ was proposed to characterize the vortex diffusion and the vortex-induced interactions [42], where $Gr$ is the Grashof number for approximating the buoyant and viscous forces of a heated fluid. In present study, we defined the nondimensional number as

$$(\alpha - 1)Gr^{1/2} = \frac{l-D}{D} \times \left[\frac{g(T_f - T_0)D^3}{\bar{T}v_A^2}\right]^{1/2} \simeq \gamma \frac{(gD)^{1/2}(l-D)}{v_A} \tag{6}$$

where $Gr$ is the ratio of buoyant forces (induced by temperature change) to viscous forces and expressed as $g(T_f - T_0)D^3/\bar{T}v^2$. In the problem of flickering flames, $Gr$ is proportional to $gD^3/v^2$, as $T_f$ is the flame temperature about 2100 $K$, $T_0$ is the ambient air temperature at 300 $K$, $\bar{T}$ is the average temperature of about 1200 $K$. Therefore, the prefactor of $\gamma = \sqrt{(T_f - T_0)/\bar{T}}$ is constant about 1.22. It should be noted that $(\alpha - 1)Gr^{1/2}$ is equivalent to a Reynolds number of $(gD)^{1/2}l/v$ that was proposed and verified in our previous work [42], where $(gD)^{1/2}$ is a characteristic velocity of the buoyancy-driven flow.

Figure 4 shows that many experiment and simulation data of dual flickering jet flames collapse to a narrow band of a transition region of $400 < (\alpha - 1)Gr^{1/2} < 500$. The frequency variation of the simulated dual flickering flames agrees well with existing literature [40, 44]. Particularly, the flickering frequency $f$ of dual flames varies with $(\alpha - 1)Gr^{1/2}$ in a nonmonotonic manner and the in-phase and anti-phase modes are distinguished by a frequency "jump", where $f/f_0$ increases from below 1 to above 1 within the transition. It should be noted that $Fr$ varies over a wide range of $O(0.001) \sim O(10)$. Consequently, the present simulation methods are verified to predict the vortex interactions of flickering flames.

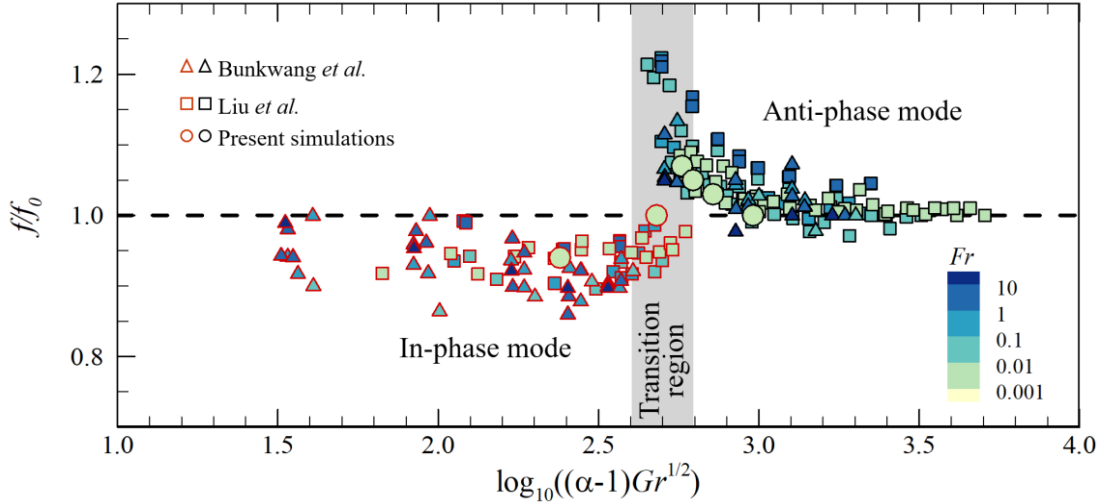

FIG. 4. Regime diagram of the dual-flame system. Data are from previous experiments [40, 44] and present simulations of dual-flame systems. The systems consist of two identical flickering diffusion flames, of which each one is characterized by the Froude number $Fr$. The nondimensional number of $(\alpha - 1)Gr^{1/2}$ is an equivalent Reynolds number [42], while the flickering frequency $f$ of dual flames is normalized by the frequency $f_0$ of single flame. A transition region of $400 < (\alpha - 1)Gr^{1/2} < 500$ distinctly separates the in- and anti-phase modes.

The physical interpretation of the bifurcation parameter $(\alpha - 1)Gr^{1/2}$ can be illustrated to make an analogy to the formation of a von Karman vortex street [42]. Eq. (6) implies that the transition between the two distinct flickering modes of the dual-flame system is dictated by the Reynolds number of $(gD)^{1/2}l/\nu$, which is defined based on the properties of the inner-side shear layers of the two flames. Besides, the flickering death mode [47, 49] is located within the transition region, in which the two flames have not been pinched off anymore and only oscillate at a small amplitude or even cease to oscillate and become steady. Yang et al. [77] interpreted that the flickering death mode can be treated as a special case of in-phase mode due to the suppression of vortex shedding at small flames that vortex interactions mainly occur far behind the flames.

### E. Stuart-Landau Model with Time-delay Coupling

The collective behaviors of coupled flickering flames are complex synchronization phenomena that widely occur across different fields such as physics, biology, chemistry, and engineering [1, 3]. In nonlinear dynamics, oscillator models are used to study the behavior of many artificial and natural systems that are collections of individuals [2]. The Stuart-Landau (S-L) model [85] has been widely used to explain the dynamical behaviors of chemical, biological, and quantum oscillator systems. The well-known model is expressed as:

$$\frac{dZ(t)}{dt} = (a + i\omega - |Z(t)|^2)Z(t) \tag{7}$$

where $Z(t) = \sqrt{a}e^{i\omega t}$ ($a > 0$) represents the oscillation state, where $\sqrt{a}$ is the oscillation amplitude and $\omega$ the oscillation frequency. The term of $|Z|^2 Z$ represents the nonlinear saturation that stabilizes the amplitude of the oscillations. If $a$ is non-positive, the system will be stable in the linear sense, and no limit cycle is formed. See Supplemental Material [83], Sec. S5.1 and Fig. S5, for more details on the illustration of the Stuart-Landau limit cycle oscillator.

In the context of coupled oscillators [85, 86], time-delay coupling has been used to produce rich and complex dynamical behaviors, such as synchronization, oscillation death, or pattern formation. In the time-delay coupling, the dynamics of an individual oscillator is influenced by delayed feedback from other individuals, introducing a time delay in their interactions. To validate the feasibility of the S-L model with time-delay coupling for the present flame oscillator problem, we took an example of dual coupled oscillators and added the interplay term of $K(Z_2(t - \tau_d) - Z_1(t))$ in the RHS of Eq. (7) for the influence of $Z_2$ oscillator on $Z_1$ oscillator, where $K$ is the coupling strength and $\tau_d$ is the time delay. See Supplemental Material [83], Sec. S5.2 and Fig. S6-S7, for more details on the numerical implementation and validation of the Stuart-Landau oscillators with the time-delay coupling.

As an illustration, we fixed the coupling strength at $K=25$ and gradually varied $\tau_d$ for the three types of interacting oscillators respectively. As shown in Fig. 5, the two S-L oscillators exhibit in-phase (no phase difference) mode in Fig. 5(a), death (very small amplitude) mode in Fig. 5(b), and anti-phase ($\pi$ phase difference) mode in Fig. 5(c) with increasing $\tau_d$, in which the transition between in-phase and anti-phase modes occurs in the range of $0.1 < \tau_d < 0.15$ (similar to the transition region of $400 < (\alpha - 1)Gr^{1/2} < 500$ for the dual flickering flames). Particularly, as shown in Fig. 5(d), the frequency trend of the present time-delay coupled S-L oscillator model is qualitatively aligned with that of the dual flame system in Fig.4. This result justifies our approach of adopting the time-delay coupled Stuart-Landau oscillators in the present problem of circular arrays of coupled flame oscillators.

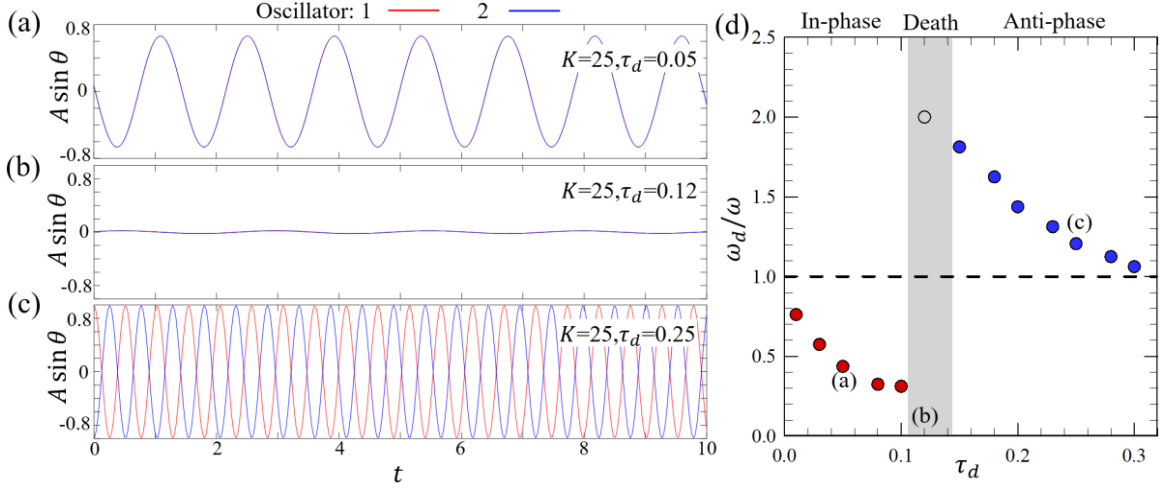

FIG. 5. The stable states of S-L oscillator 1 and 2 in (a) in-phase mode, (b) death mode, (c) anti-phase mode as $K=25$, and $\tau_d = 0.05$, 0.12, and 0.25, respectively. The oscillator's real parts of $A_1(t)\sin\theta_1(t)$ and $A_2(t)\sin\theta_2(t)$ are plotted in red and blue lines respectively, where $A$ is the amplitude and $\theta$ is the phase angle. As discussed in Sec. S4.1, $a=1$ and $\omega=10$ are used to keep each uncoupled oscillator in the limit cycle. It should be noted that the death mode is defined as the oscillation of $A < 0.1$ in the present study. (d) Three dynamical modes of two identical S-L oscillators with time-delay coupling in the parameter space of the normalized oscillation frequency $\omega_d/\omega$ and time delay $\tau_d$.

As illustrated in Sec. III.D, the flame mode significantly depends on the vortex interactions in the gap flow between adjacent flames [42]. In this study, we considered the oscillator's local interplay for the interactions among octuple flames in a circle array in Fig. 3, where each oscillator is coupled only to its nearest neighbors. Consequently, the nearest neighbor coupling is utilized for approximately solving the complex coupling problem. For multiple interacting bodies distributed uniformly along a ring, the extension of Eq. (7) can be approximated to just two coupling terms, caused by nearest neighbor oscillators. Hence, the S-L model for $Z_j$, one of the $j = 1\sim 8$ oscillators, has the form

$$\frac{dZ_j}{dt} = \left(a + \omega i - |Z_j|^2\right)Z_j + K\left(Z_{j+1}(t-\tau_d) - Z_j(t)\right) + K\left(Z_{j-1}(t-\tau_d) - Z_j(t)\right) \quad (8)$$

where all oscillators are identical with $a=1$ and $\omega=10$ for simplification [38] and arranged in a closed ring with the same interval, forming a local coupling unite of $j-1$, $j$, and $j+1$ oscillators with periodic boundary conditions for the circulant coupling, particuarlly, the unite 7-8-9(1) for $Z_8$ oscillator and the unite 0(8)-1-2 for $Z_1$ oscillator. Similarly, Biju et al. [46] employed periodic limits to enforce the ring topology of an oscillator network. It is noteworthy that the nearest neighbor coupling is a local model, unlike global and nonlocal coupling models [85], where the coupling has a global dependence on the geometrical arrangement of the oscillators.

## IV. RESULTS AND DISCUSSIONS

### A. Computational Identification of Dynamical Modes

The collective behaviors of the octuple flame system, consisting of eight identical flame individuals at $Re = 100$ and $Fr = 0.28$ (the flickering frequency $f_0 = 10$ Hz), were investigated by varying the dimensionless flame-gap distance $\alpha$ from 1.6 to 4.5. These ten cases of octuple-flickering flames with $(\alpha - 1)Gr^{1/2} = 152 - 885$ cover the range of coupling effect of the dual flame system, as discussed in Sec. III.D. We found five distinct flame modes and three dynamical regimes for collective behaviors of present circular flame arrays by characterizing their flickering amplitude and phase difference. Table III shows the detailed parameters and classifications of the simulated circular arrays of flame oscillators. For a larger circular array of present flickering flames, the adjacent flames would be weak and render a case of no interest to the present study.

The relevant observations for those modes can be made as follows: 1) the merged (hereinafter referred to as Me) mode indicates that the octuple flames can merge into a bigger one, as they are too close to each other (e.g., $\alpha =$1.6 and 2.4); 2) the in-phase (hereinafter referred to as IP) mode appears as these flames are sufficiently separated but can flicker synchronously without a phase difference (e.g., $\alpha =$2.8, 3.0, and 3.1); 3) the rotation and flickering death (hereinafter referred to as RFD) mode appears as the flames alternatively flicker in the azimuthal direction and the flame flicker is suppressed (e.g., $\alpha =$3.4); 4) the partially flickering death mode (hereinafter referred to as PFD) occurs when the octuple can be divided into several subgroups (e.g., the 1-2-3 flames), behaving behaves like the partially flickering death mode of triple flame system (e.g., $\alpha =$3.5 and 3.8); 5) the anti-phase (hereinafter referred to as AP) mode is that the eight flames are divided into two groups (1-3-5-7 flames and 2-4-6-8 flames) with a constant phase difference of $\pi$ (e.g., $\alpha$ =4.0 and 4.5). In terms of flickering amplitude and phase difference, the five flame modes can be sorted into three regimes: in-phase regime (i.e., Me and IP modes without any phase difference among all flames), flickering suppression regime (i.e., RFD and PFD modes with amplitude suppression of flames, at least one), and anti-phase regime (i.e., AP mode with $\pi$ phase difference for two clusters). The details of each dynamic mode will be elaborated in later sections.

TABLE III. Main information of simulation cases.

| No. | $\alpha$ | $(\alpha-1)Gr^{1/2}$ | Flame mode | Dynamical regime |
|---|---|---|---|---|
| 1 | 1.6 | 152 | Merged (Me) mode: all flames are merged into a big flame. | |
| 2 | 2.4 | 354 | | In-phase regime (no phase difference) |
| 3 | 2.8 | 455 | In-phase (IP) mode: all flames flicker as one group without phase difference. | |
| 4 | 3.0 | 506 | | |
| 5 | 3.1 | 531 | | |
| 6 | 3.4 | 607 | Rotation and flickering death (RFD) mode: all flames are not pinched off and alternatively oscillate in the azimuthal direction. | Flickering suppression regime (amplitude suppression) |
| 7 | 3.5 | 632 | Partially flickering death (PFD) mode: the subgroup of flame 1-2-3 behaves the same as the partially flickering death mode of the triple flame system. | |
| 8 | 3.8 | 708 | | |
| 9 | 4.0 | 759 | Anti-phase (AP) mode: eight flames are divided into two groups (1-3-5-7 and 2-4-6-8) with a constant phase difference of $\pi$. | Anti-phase regime ($\pi$ phase difference) |
| 10 | 4.5 | 885 | | |

Furthermore, the flame frequencies of five representative cases of these distinct modes are calculated from the flame time series, for example, the velocity magnitude at the center of each nozzle, as shown in Fig. 6(a)-10(a). The same flame frequencies are calculated based on the velocity magnitude at the $3D$ downstream of each nozzle (see Fig. S8 in Supplemental Material [83]). Compared to the self-excited motion of a single flame, collective flames oscillate at different frequencies. Particularly, due to the three-dimensional complexity of the problem, we adopted the time-varying flame quantity in the transverse section ($X-Y$ plane) to characterize the collective behaviors, which can be effectively depicted by the change of the flame temperature in the horizontal plane, as illustrated in Sec. 3.1.

**Merged (Me) mode**: Figure 6 shows the representative case $\alpha = 2.4$ for the merged mode of octuple flickering flames. During the periodic process of $\tau = 227$ ms, the eight flames can be well recognized according to the time-varying temperature distributions in the horizontal plane of $Z/D=6.0$, as shown in Fig. 7(b). Particularly, their high temperature ($\geq 2000\ K$) can be connected from $2\tau/6$ to $3\tau/6$. The connected zone indicates that the flame sheets are merged and subsequently form a bigger flame surface.

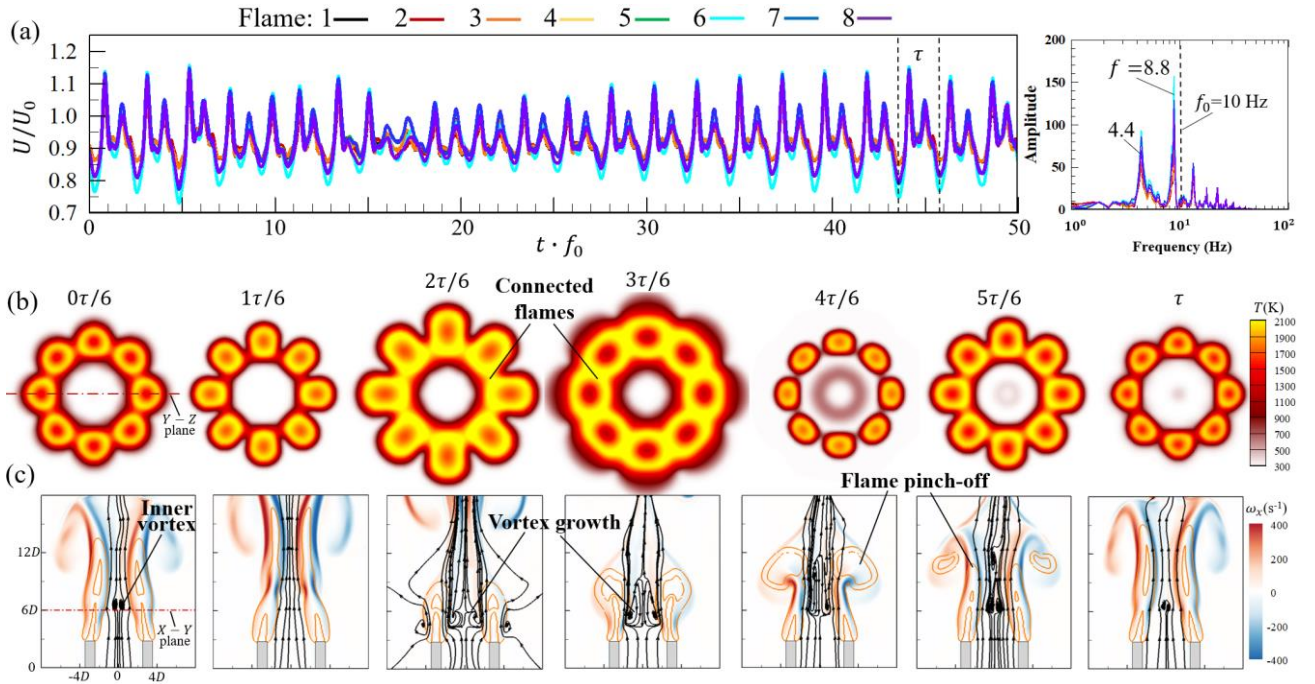

FIG. 6. The periodic collective process of the circular array of octuple flickering flames for $\alpha = 2.4$: (a) time and frequency domains of the velocity magnitudes at the center of each nozzle, (b) horizontal $X - Y$ plane at $Z = 6D$, and (c) longitudinal $Y - Z$ plane at $X = 0D$.

Fig. 6(c) corresponds to the dynamical behaviors of two flames (flames 3 and 7) in the longitudinal plane. An inner vortex, denoted by curled streamlines, appears inside the flames. The size and location of the recirculation zone evolve with the flames. It should be noted that the vortex grows up when flames are lengthened and deformed. At $5\tau/6$, the flame is pinched off and the vortex shedding occurs. The merged flames flicker at a primary frequency of 8.8 Hz, smaller than that of a single flickering flame. Also, we observed a lower frequency of 4.4 Hz (called the secondary frequency) in Fig. 6(a) and conjectured it results from the emergence of the inner vortex, as the vortex has about twice the length scale of the outside vortex. Interestingly, the time-varying velocity magnitude waves as an "M" shape with unequal left and right peaks.

**In-phase (IP) mode**: the collective behaviors of eight flames in the case $\alpha =2.8$ occur in a flickering way, as shown in Fig. 7. As there is no connection between the high-temperature zones in Fig. 7(b), these flames are not merged but oscillate individually with no phase difference. Fig. 7(c) clearly shows that the collective flames are affected by the inner vortex and the flame dynamics are closely associated with the vortex evolution. These synchronized flames exhibit the up-and-down and the back-and-forth deformation, appearing like a "worship" [37]. The vortex reconnection plays an essential role in the collective dynamics, as similarly illustrated in the triple-flame system in our previous work [77]. Besides, their primary and secondary frequencies, 8.4 Hz and 4.2 Hz shown in Fig. 7(a) respectively, are slightly smaller than those of the case $\alpha =2.4$. This is due to the bigger diameter of the circular array.

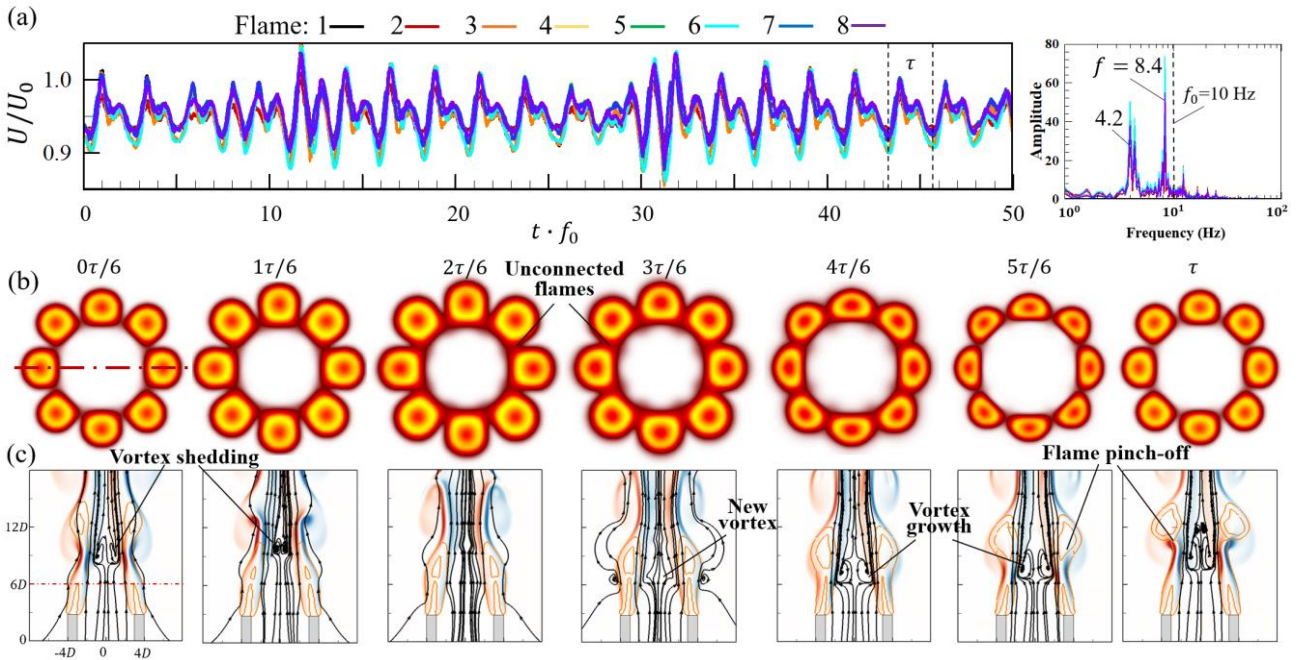

FIG. 7. The periodic collective process of a circular array of octuple flickering flames for $\alpha =2.8$: (a) time and frequency domains of the velocity magnitudes at the center of each nozzle, (b) horizontal $X-Y$ plane at $Z=6D$, and (c) longitudinal $Y-Z$ plane at $X=0D$.

**Rotation and flickering death (RFD) mode**: In the case $\alpha =3.4$ shown in Fig. 8, the flame circular array becomes bigger and the collective behaviors vary significantly. These flames have an azimuthal motion and the pinch-off of each flame vanishes. As shown in Fig. 8(b), the transverse temperature shapes of each two diagonal flames deform in opposite directions (denoted by the opposite arrows). The collective flames regularly form the circumferential motion (see Supplemental Material [83], Fig. S9, for the height variation of flames on the cylindrical slice. More specifically, the inserted figure shows the three-dimensional pattern of flame sheets (denoted by the iso-surface of the heat release rate). A twisting deformation of flames in the anticlockwise direction can be seen. In addition, we noted that the eight flames have no pinch-off during the periodic process of $\tau = 179$ ms and the inner vortex occurs near the top part of flames, as shown in Fig. 8(c). In fact, the primary frequency of $f=12.0$ Hz shown in Fig. 8(a) could characterize the swaying of the flame head, instead of the vertical oscillation. This collective behavior is similar to the flickering death mode of a dual- or triple-flame system [47, 49], in which the flicker of one flame is suppressed by the vortex-induced flow of the adjacent flames [77].

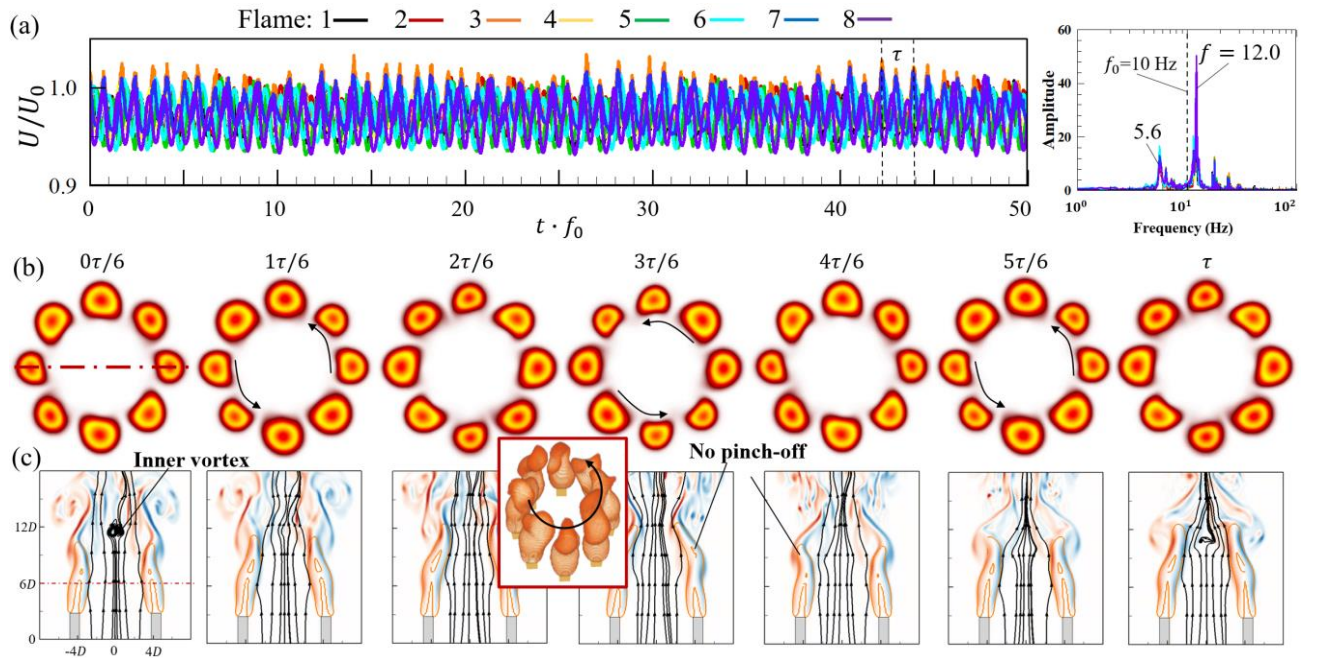

FIG. 8. The periodic collective process of a circular array of octuple flickering flames for $\alpha = 3.4$: (a) time and frequency domains of the velocity magnitudes at the center of each nozzle, (b) horizontal $X - Y$ plane at $Z = 6D$, and (c) longitudinal $Y - Z$ plane at $X = 0D$.

**Partially flickering death (PFD) mode**: In the case $\alpha = 3.8$, the longitudinal diagonal flames

(flames 1 and 5) flicker in an in-phase way and so do the horizontal pair of flames 3 and 7. However, these two pairs have a $\pi$ phase difference, as shown in Fig. 9(b). Meanwhile, the other flames (flames 2, 4, 6, and 8) are in the flickering death mode. For the collective flicker of 1-3-5-7 flames, its primary frequency is $f=11.8$ Hz (about a period of $\tau=85$ ms) in Fig. 9(a). Figure 9 (c) shows that the inner vortex forms far away from the flames, so the negligible influence of the downstream vortex on the flames may cause no secondary frequency.

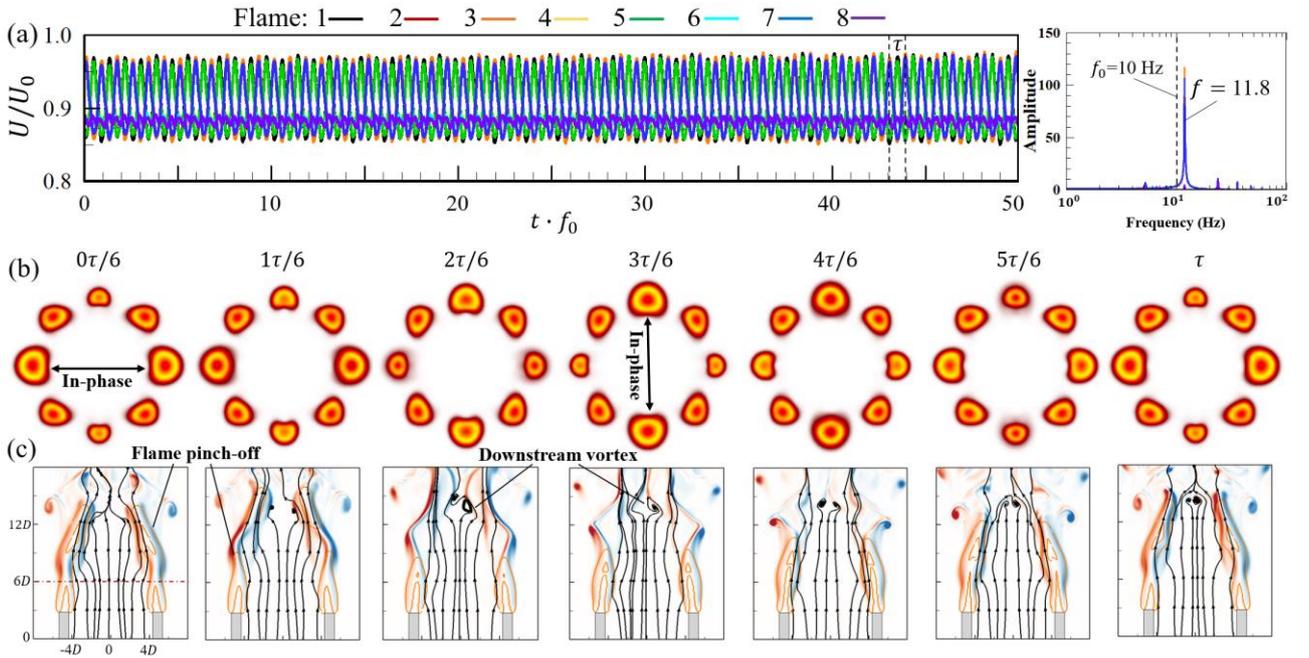

FIG. 9. The periodic collective process of a circular array of octuple flickering flames for $\alpha =3.8$: (a) time and frequency domains of the velocity magnitudes at the center of each nozzle, (b) horizontal $X-Y$ plane at $Z=6D$, and (c) longitudinal $Y-Z$ plane at $X=0D$.

**Anti-phase (AP) mode**: When eight flames are relatively away from each other, for example in the case $\alpha =4.0$, we can see the flames are separated into two groups of 1-3-5-7 flames and 2-4-6-8 flames in Fig. 10(b). In each group, the in-phase mode appears as the four flames flicker synchronously with no phase difference. The two group flames alternatively flicker with a phase difference of $\pi$. The collective behavior of octuple flickering flames is similar to the anti-phase mode of the dual flame system [42]. Besides, we noted that the inner vortex forms near the flame downstream in Fig. 10(c). Thus, the shear layers of adjacent flames in the circular array, exhibiting a distinct asymmetric feature, play an important role in the collective behavior. In this case, the collective motion only has a primary frequency of 12.4 Hz in Fig. 10(a), of which the frequency increase (2.4 Hz higher than $f_0$) is consistent with the finding of the anti-phase mode of the dual flame system, as mentioned in Sec. III.D.

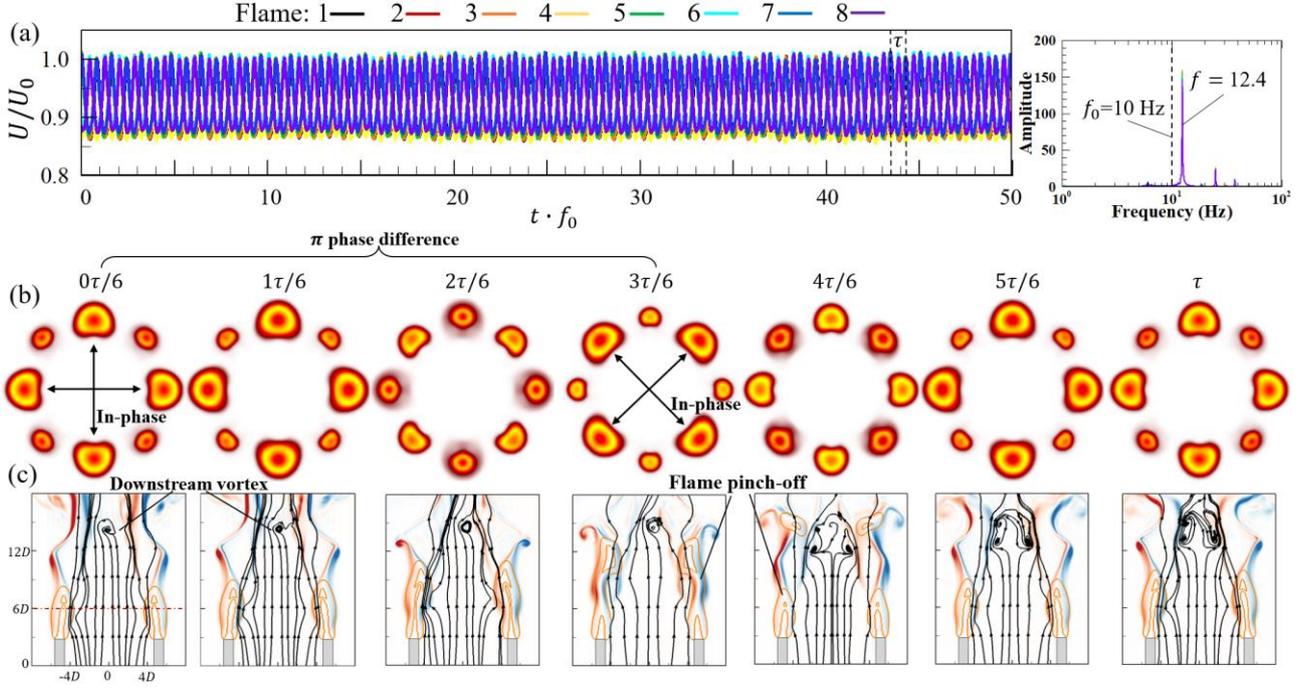

FIG. 10. The periodic collective process of a circular array of octuple flickering flames for $\alpha = 4.0$: (a) time and frequency domains of the velocity magnitudes at the center of each nozzle, (b) horizontal $X - Y$ plane at $Z = 6D$, and (c) longitudinal $Y - Z$ plane at $X = 0D$.

**B. Parametric Transition of Dynamical Modes**

The non-dimensional frequencies, $f/f_0$, of circular arrays of octuple flickering flames are plotted against $(\alpha - 1)Gr^{1/2}$ are shown in Fig. 11(a). This frequency variation trend shows a bifurcation transition (denoted by the grey zone of $(\alpha - 1)Gr^{1/2} = 655 \pm 55$) between the in-phase mode (a lower-frequency collective state with secondary frequency) and the anti-phase mode (a higher-frequency collective state without secondary frequency). It should be noted that the collective effect causes the transition range of about $600{\sim}710$ to be higher than $400{\sim}500$ of the dual flame system in Fig. 4.

The transition covers the RFD and PFD modes, in which the flickering amplitudes of collective flames are suppressed neither completely or partially. Previous experimental observations [47, 49] reported similar phenomena of amplitude death in combustion systems with a few flames (≤ 3). In fact, the death state occurs commonly in various dynamical systems, such as neuron and brain cells and prey-predator systems. Particularly, nonflickering flame mode may reduce the noise generation reduced in a combustion chamber. Different from the moving-burner technique to achieve steady-state combustion [87], the present study gives a feasible way to design flame layouts in a circular array system for flicker suppression.

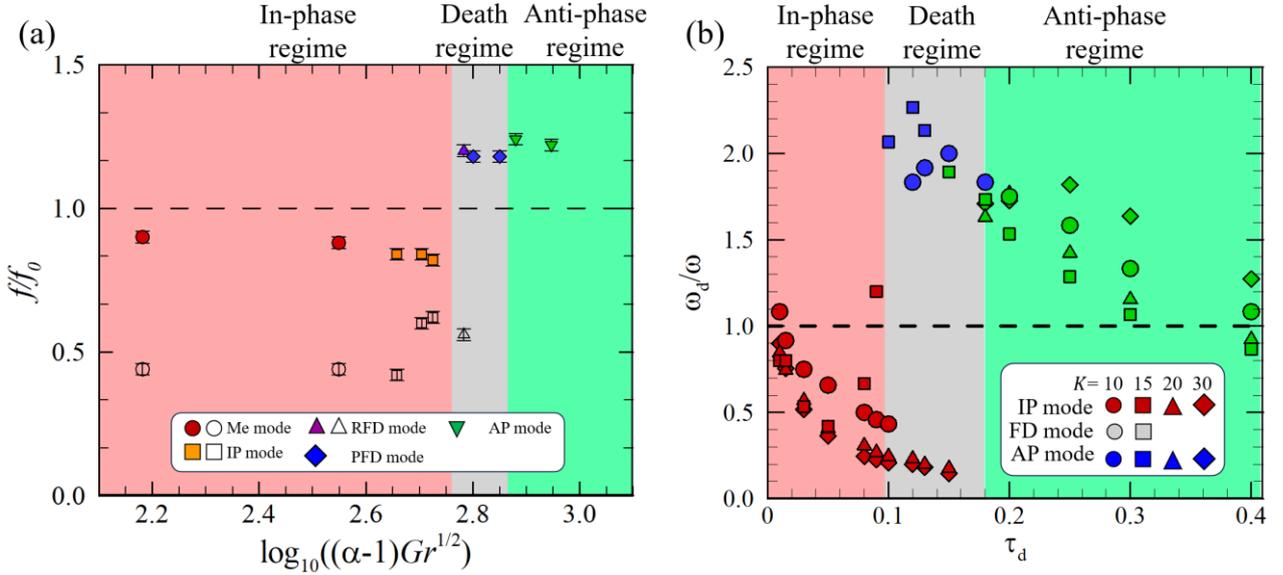

FIG. 11. Regime diagram of (a) octuple flame oscillators and (b) octuple identical S-L oscillators with time-delay coupling. In the flame system, the eight identical flickering diffusion flames exhibit five distinct modes with increasing the nondimensional number of $(\alpha - 1)Gr^{1/2}$ [42]. Their frequencies $f$ are normalized by the natural frequency $f_0$ and small $(\alpha - 1)Gr^{1/2}$ cases, including the merged, in-phase, rotation & flickering death modes, have secondary frequencies. A transition region of $600 < (\alpha - 1)Gr^{1/2} < 710$, where the flicker death occurs in some or all flames, distinctly separates the in- and anti-phase regimes. In the S-L system, using the nearest neighbor coupling for the oscillator interactions, Eq. (11) is solved to obtain three modes and their $\omega_d/\omega$ at different $K$ and $\tau_d$. The collective states of flames are identified according to the phase difference and amplitude, as mentioned in Sec. III C. A transition region of $0.01 < \tau_d < 0.4$ distinctly separates the in- and anti-phase regimes.

### C. Stuart-Landau Oscillators with Nearest Neighbor Coupling

In the present study, Eq. (8) is used as a toy model to study the collective states of multiple flame oscillators. Figure 11(b) shows the results of the present model within $10 < K < 30$ and $0.01 < \tau_d < 0.4$ for octuple S-L oscillator systems. In the regime diagram, the in-phase, death, and anti-phase modes are reproduced with increasing $\tau_d$, while the death mode is absent when $K$ is relatively large. In general, the frequency trend of $\omega_d/\omega$ has a jump variation within the transition region of $0.10 < \tau_d < 0.18$. Like the dual oscillator system, the eight oscillators in the in-phase state have lower frequencies than the natural frequency, while the frequencies of those in the anti-phase state are higher. These results are consistent with the observations on the three regimes in the octuple flickering flame systems in Fig. 11(a). Specially, the small $\tau_d = 0\sim0.1$ region roughly corresponds to the region of merged and in-phase states, where $(\alpha - 1)Gr^{1/2}$ is smaller than 600; the situation that the death mode occurs at the appropriate $\tau_d$ and $K$ is similar to those cases of rotation & death and partial death modes in a narrow region of $(\alpha - 1)Gr^{1/2} = 600 - 710$; a larger $\tau_d$ plays a similar role as a larger $(\alpha - 1)Gr^{1/2}$ to divide the eight oscillators into two anti-phase clusters (i.e., four oscillators are in-phase in each cluster but anti-phase with the other four). Therefore, the S-L model used here has brought out a great capability of reproducing the general features and collective modes of flickering flame in circular arrays.

It is noted that Eq. (8) did not reproduce a few modes, such as all flames in the merged state, all flames possessing an azimuthal motion, and partial flames in the death state. The limitation of the S-L model may be due to two causes. First, the S-L model is represented by a limit cycle oscillator, consisting of amplitude and frequency, while the flame is a high-dimensional system, in which spatiotemporal interactions could be multimodal. Increasing flame numbers or varying flame layouts is likely to make the flame and vortex interactions complicated, thereby forming distinct dynamical behaviors from the collective perspective. Second, the nearest neighbor coupling of S-L oscillators is an approximate model to mimic the interaction between adjacent flames. Collective states and essential characteristics of oscillators apparently rely on coupling scenarios, for example, all-to-all (global) coupling and spatially varying (non-local) coupling. The additional dimension is required for the global or nonlocal coupled terms and accordingly introduces new equilibrium states in the system such as traveling waves and in higher dimensions spiral patterns or scroll waves [85]. The influences of the global coupling on the collective behaviors of multiple oscillators merit future work.

## V. CONCLUDING REMARKS

In nature, local interactions between individual entities often give rise to patterns or collective behaviors in groups. The flickering flame, as a flame oscillator, has been used widely in recent years to study the nonlinear dynamics of complex systems. In this study, we computationally and theoretically investigated the emergence of synchronization events in circular arrays of octuple identical flame oscillators (i.e., small-scale flickering buoyant diffusion flames of $Fr = 0.28$). Parametric studies from the perspectives of flickering flame simulation and S-L oscillator model were carried out in the range of a bifurcation parameter of $(\alpha - 1)Gr^{1/2} = 100 \sim 900$ and in the two-parameter space of $10 < K < 30$ and $0.01 < \tau_d < 0.4$.

In the present simulations, five distinct dynamical modes were identified by comparing their frequencies and phase differences and were classified into three regimes, namely the in-phase regime including the merged mode (all flames are merged into a big one with a lower flickering frequency), in which an inner vortex forms in the circle array and coexists with the outer buoyance-induced toroidal vortex, and the in-phase mode (all separated flame individuals flicker synchronously without phase difference), in which the vortex reconnection keeps all flames in same manner; the flickering death regime including the rotation & flickering death mode (the flicker of all flames is suppressed and they collectively form an azimuthal motion), in which the flame flicker is suppressed by the vortex-induced flow of other flames, and the partially flickering death mode (a part of the eight identical flames is flickering death and the rest flames flicker in an in-phase or anti-phase synchronization), in which the inner vortex occurs far away from the flames and the vortex-induced flow causes the flickering death of some flames; and the anti-phase regime including the anti-phase mode (flames are separated into two groups with a $\pi$ phase difference, but flames flicker synchronously without phase difference in each group), in which the shear layers of adjacent flames exhibit a distinct asymmetric feature.

Understanding transitions between various dynamical modes is of theoretical and practical importance. We found a bifurcation transition of $(\alpha - 1)Gr^{1/2} = 620 \pm 50$ between the in-phase mode and the anti-phase mode in the regime diagram of flame circular arrays. The transition separates lower-frequency collective states with a secondary frequency and higher-frequency collective states without a secondary frequency. Those secondary frequencies could be due to the formation of a bigger inner vortex in circle arrays. In the transition, the collective effect can suppress the flame flicker totally or partially (i.e., the flickering death). Thus, designing a circular array system with a suitable $(\alpha - 1)Gr^{1/2}$ is a feasible way to generate a death mode, thereby reducing noise generation and possibly relieving thermoacoustic oscillations [51] that usually occur in annular combustors.

Three distinct states including the in-phase, death, and anti-phase modes were theoretically reproduced by using the S-L model with a time-delay coupling. Within $10 < K < 30$ and $0.01 < \tau_d < 0.4$, octuple S-L oscillators exhibit the in-phase, death, and anti-phase modes, while the death mode is absent when $K$ is relatively large. However, the merged, rotation, and partial death modes were not reproduced by the model. Considering the S-L model with a time-delay coupling can reproduce all the modes of the dual flame system, we attributed the limitation of the S-L model to the lack of higher-dimensional representation of flame dynamics and the absence of nonlocal (or even global) coupling. More sophisticated S-L modelling of the present system merits future work.

Nevertheless, the present study provides a novel approach to explore synchronization in complex collective features of flames in circular arrays. The physical modelling based on vortex-dynamics mechanisms and its relation with the toy model of Stuart-Landau oscillators remain a problem to be solved in the future. Extending the present study to include the influences of turbulent flow, wall confinement, and combustion-acoustics interaction [11, 88] in larger-size annular combustion systems merits future works.

# ACKNOWLEDGMENT

This work is supported by the National Natural Science Foundation of China (No. 52176134) and partially by the APRC-CityU New Research Initiatives/Infrastructure Support from Central of City University of Hong Kong (No. 9610601). The authors also acknowledge the National

# REFERENCES

[1] A. Pikovsky, M. Rosenblum, J. Kurths, Synchronization: a universal concept in nonlinear science, American Association of Physics Teachers, 2002.
[2] D.A. Paley, N.E. Leonard, R. Sepulchre, D. Grunbaum, J.K. Parrish, Oscillator models and collective motion, IEEE Control Systems Magazine 27 (2007) 89-105.
[3] S.H. Strogatz, Nonlinear dynamics and chaos: with applications to physics, biology, chemistry, and engineering, CRC Press2018.
[4] G. Staffelbach, L. Gicquel, G. Boudier, T. Poinsot, Large Eddy Simulation of self excited azimuthal modes in annular combustors, Proc. Combust. Inst. 32 (2009) 2909-2916.
[5] G. Ghirardo, M.R. Bothien, Quaternion structure of azimuthal instabilities, Phys. Rev. Fluids 3 (2018) 113202.
[6] G. Vignat, D. Durox, T. Schuller, S. Candel, Combustion dynamics of annular systems, Combust. Sci. Technol. 192 (2020) 1358-1388.
[7] A. Faure-Beaulieu, T. Indlekofer, J.R. Dawson, N. Noiray, Imperfect symmetry of real annular combustors: beating thermoacoustic modes and heteroclinic orbits, J. Fluid Mech. 925 (2021).
[8] H.T. Nygård, G. Ghirardo, N.A. Worth, Azimuthal flame response and symmetry breaking in a forced annular combustor, Combust. Flame 233 (2021) 111565.
[9] N. Peake, A.B. Parry, Modern challenges facing turbomachinery aeroacoustics, Annu. Rev. Fluid Mech. 44 (2012) 227-248.
[10] K. Moon, Y. Choi, K.T. Kim, Experimental investigation of lean-premixed hydrogen combustion instabilities in a can-annular combustion system, Combust. Flame 235 (2022) 111697.
[11] P.E. Buschmann, N.A. Worth, J.P. Moeck, Thermoacoustic oscillations in a can-annular model combustor with asymmetries in the can-to-can coupling, Proc. Combust. Inst. 39 (2023) 5707-5715.
[12] X. Liu, D. Zhao, D. Guan, S. Becker, D. Sun, X. Sun, Development and progress in aeroacoustic noise reduction on turbofan aeroengines, Prog. Aerosp. Sci. 130 (2022) 100796.
[13] D.S. Chamberlin, A. Rose, The flicker of luminous flames, Proc. Combust. Inst. 1-2 (1948) 27-32.
[14] D. Durox, T. Yuan, F. Baillot, J. Most, Premixed and diffusion flames in a centrifuge, Combust. Flame 102 (1995) 501-511.
[15] J.-M. Most, P. Mandin, J. Chen, P. Joulain, D. Durox, A.C. Fernande-Pello. Influence of gravity and pressure on pool fire-type diffusion flames. In: editor^editors. Symposium (International) on Combustion; 1996: Elsevier. p. 1311-1317.
[16] D. Durox, T. Yuan, E. Villermaux, The effect of buoyancy on flickering in diffusion flames, Combust. Sci. Technol. 124 (1997) 277-294.
[17] X. Jiang, K. Luo, Combustion-induced buoyancy effects of an axisymmetric reactive plume, Proc. Combust. Inst. 28 (2000) 1989-1995.
[18] H. Sato, K. Amagai, M. Arai, Diffusion flames and their flickering motions related with Froude numbers under various gravity levels, Combust. Flame 123 (2000) 107-118.
[19] J. Carpio, M. Sánchez-Sanz, E. Fernández-Tarrazo, Pinch-off in forced and non-forced, buoyant laminar jet diffusion flames, Combust. Flame 159 (2012) 161-169.
[20] D. Moreno-Boza, W. Coenen, A. Sevilla, J. Carpio, A. Sánchez, A. Liñán, Diffusion-flame flickering as a hydrodynamic global mode, J. Fluid Mech. 798 (2016) 997-1014.
[21] X. Xia, P. Zhang, A vortex-dynamical scaling theory for flickering buoyant diffusion flames, J. Fluid Mech. 855 (2018) 1156-1169.
[22] G.M. Nayak, P. Kolhe, S. Balusamy, Role of buoyancy induced vortices in a coupled-mode of oscillation in laminar and turbulent jet diffusion flames, Flow, Turbulence and Combustion, (2022)

1-19.

[23] Y. Araya, H. Ito, H. Kitahata, Bifurcation structure of the flame oscillation, Physical Review E 105 (2022) 044208.

[24] L.-D. Chen, J. Seaba, W. Roquemore, L. Goss, Buoyant diffusion flames, Proc. Combust. Inst. 22 (1989) 677-684.

[25] K.S. Pasumarthi, A.K. Agrawal, Schlieren measurements and analysis of concentration field in self-excited helium jets, Phys. Fluids 15 (2003) 3683-3692.

[26] N.T. Wimer, C. Lapointe, J.D. Christopher, S.P. Nigam, T.R. Hayden, A. Upadhye, M. Strobel, G.B. Rieker, P.E. Hamlington, Scaling of the puffing Strouhal number for buoyant jets and plumes, J. Fluid Mech. 895 (2020) A26.

[27] Y. Zhang, Y. Yang, Y. Wei, S. Liu, Vortex shedding controlled combustion of the wake flame of an n-heptane wetted porous sphere, AIP Adv. 12 (2022) 105216.

[28] S. Thirumalaikumaran, G. Vadlamudi, S. Basu, Insight into flickering/shedding in buoyant droplet-diffusion flame during interaction with vortex, Combust. Flame 240 (2022) 112002.

[29] N. Fujisawa, Y. Matsumoto, T. Yamagata, Influence of co-flow on flickering diffusion flame, Flow, Turbulence and Combustion 97 (2016) 931-950.

[30] T. Yang, Y. Ma, P. Zhang, Dynamical behavior of small-scale buoyant diffusion flames in externally swirling flows, Symmetry 16 (2024) 292.

[31] V.R. Katta, W.M. Roquemore, A. Menon, S.-Y. Lee, R.J. Santoro, T.A. Litzinger, Impact of soot on flame flicker, Proc. Combust. Inst. 32 (2009) 1343-1350.

[32] G. Legros, T. Gomez, M. Fessard, T. Gouache, T. Ader, P. Guibert, P. Sagaut, J. Torero, Magnetically induced flame flickering, Proc. Combust. Inst. 33 (2011) 1095-1103.

[33] Y. Xiong, S.H. Chung, M.S. Cha, Instability and electrical response of small laminar coflow diffusion flames under AC electric fields: Toroidal vortex formation and oscillating and spinning flames, Proc. Combust. Inst. 36 (2017) 1621-1628.

[34] M. Ahn, D. Lim, T. Kim, Y. Yoon, Pinch-off process of Burke–Schumann flame under acoustic excitation, Combust. Flame 231 (2021) 111478.

[35] T. Chen, X. Guo, J. Jia, J. Xiao, Frequency and phase characteristics of candle flame oscillation, Sci. Rep. 9 (2019) 1-13.

[36] K. Okamoto, A. Kijima, Y. Umeno, H. Shima, Synchronization in flickering of three-coupled candle flames, Sci. Rep. 6 (2016) 1-10.

[37] D.M. Forrester, Arrays of coupled chemical oscillators, Sci. Rep. 5 (2015) 1-7.

[38] K. Manoj, S.A. Pawar, S. Dange, S. Mondal, R. Sujith, E. Surovyatkina, J. Kurths, Synchronization route to weak chimera in four candle-flame oscillators, Physical Review E 100 (2019) 062204.

[39] P.W. Anderson, More Is Different: Broken symmetry and the nature of the hierarchical structure of science, Science 177 (1972) 393-396.

[40] A. Bunkwang, T. Matsuoka, Y. Nakamura, Similarity of dynamic behavior of buoyant single and twin jet-flame (s), Journal of Thermal Science and Technology 15 (2020) 1-14.

[41] T. Tokami, M. Toyoda, T. Miyano, I.T. Tokuda, H. Gotoda, Effect of gravity on synchronization of two coupled buoyancy-induced turbulent flames, Physical Review E 104 (2021) 024218.

[42] T. Yang, X. Xia, P. Zhang, Vortex-dynamical interpretation of anti-phase and in-phase flickering of dual buoyant diffusion flames, Phys. Rev. Fluids 4 (2019) 053202.

[43] H. Kitahata, J. Taguchi, M. Nagayama, T. Sakurai, Y. Ikura, A. Osa, Y. Sumino, M. Tanaka, E. Yokoyama, H. Miike, Oscillation and synchronization in the combustion of candles, The Journal of Physical Chemistry A 113 (2009) 8164-8168.

[44] L. Changchun, L. Xinlei, G. Hong, D. Jun, Z. Shasha, W. Xueyao, C. Fangming, On the influence of distance between two jets on flickering diffusion flames, Combust. Flame 201 (2019) 23-30.

[45] K. Manoj, S.A. Pawar, R. Sujith, Experimental investigation on the susceptibility of minimal networks to a change in topology and number of oscillators, Physical Review E 103 (2021) 022207.

[46] A.E. Biju, S. Srikanth, K. Manoj, S.A. Pawar, R. Sujith, Dynamics of minimal networks of limit cycle oscillators, Nonlinear Dynamics, (2024) 1-20.
[47] Y. Chi, T. Yang, P. Zhang, Dynamical mode recognition of triple flickering buoyant diffusion flames in Wasserstein space, Combust. Flame 248 (2023) 112526.
[48] A. Gergely, B. Sándor, C. Paizs, R. Tötös, Z. Néda, Flickering candle flames and their collective behavior, Sci. Rep. 10 (2020) 1-13.
[49] K. Manoj, S.A. Pawar, R. Sujith, Experimental evidence of amplitude death and phase-flip bifurcation between in-phase and anti-phase synchronization, Sci. Rep. 8 (2018) 1-7.
[50] Y. Chi, Z. Hu, T. Yang, P. Zhang, Synchronization modes of triple flickering buoyant diffusion flames: Experimental identification and model interpretation, Physical Review E 109 (2024) 024211.
[51] M.P. Juniper, R.I. Sujith, Sensitivity and nonlinearity of thermoacoustic oscillations, Annu. Rev. Fluid Mech. 50 (2018) 661-689.
[52] J.-F. Parmentier, P. Salas, P. Wolf, G. Staffelbach, F. Nicoud, T. Poinsot, A simple analytical model to study and control azimuthal instabilities in annular combustion chambers, Combust. Flame 159 (2012) 2374-2387.
[53] M. Bauerheim, J.-F. Parmentier, P. Salas, F. Nicoud, T. Poinsot, An analytical model for azimuthal thermoacoustic modes in an annular chamber fed by an annular plenum, Combust. Flame 161 (2014) 1374-1389.
[54] J.P. Moeck, D. Durox, T. Schuller, S. Candel, Nonlinear thermoacoustic mode synchronization in annular combustors, Proc. Combust. Inst. 37 (2019) 5343-5350.
[55] J.G. Aguilar, J. Dawson, T. Schuller, D. Durox, K. Prieur, S. Candel, Locking of azimuthal modes by breaking the symmetry in annular combustors, Combust. Flame 234 (2021) 111639.
[56] B. Ahn, H.T. Nygård, N.A. Worth, Z. Yang, L.K. Li, Longitudinal and azimuthal thermoacoustic modes in a pressurized annular combustor with bluff-body-stabilized methane-hydrogen flames, Phys. Rev. Fluids 9 (2024) 053907.
[57] M. Bauerheim, F. Nicoud, T. Poinsot, Progress in analytical methods to predict and control azimuthal combustion instability modes in annular chambers, Phys. Fluids 28 (2016) 021303.
[58] P.E. Buschmann, G.A. Mensah, J.P. Moeck, Intrinsic thermoacoustic modes in an annular combustion chamber, Combust. Flame 214 (2020) 251-262.
[59] G. Bonciolini, A. Faure-Beaulieu, C. Bourquard, N. Noiray, Low order modelling of thermoacoustic instabilities and intermittency: flame response delay and nonlinearity, Combust. Flame 226 (2021) 396-411.
[60] Y. Weng, V.R. Unni, R. Sujith, A. Saha, Synchronization-based model for turbulent thermoacoustic systems, Nonlinear Dynamics 111 (2023) 12113-12126.
[61] Turbulent jet flame database generated for sooty fuels. https://crf.sandia.gov/turbulent-jet-flame-database-generated-for-sooty-fuels/ (accessed December 2023).
[62] K. Sahu, A. Kundu, R. Ganguly, A. Datta, Effects of fuel type and equivalence ratios on the flickering of triple flames, Combust. Flame 156 (2009) 484-493.
[63] K. McGrattan, S. Hostikka, R. McDermott, J. Floyd, C. Weinschenk, K. Overholt, Fire dynamics simulator user's guide, NIST Special Publication 1019 (2013) 1-339.
[64] R.G. Rehm, H.R. Baum, The equations of motion for thermally driven, buoyant flows, Journal of research of the National Bureau of Standards 83 (1978) 297.
[65] K. McGrattan, R. Rehm, H. Baum, Fire-driven flows in enclosures, Journal of Computational Physics 110 (1994) 285-291.
[66] H.R. Baum, O. Ezekoye, K.B. McGrattan, R.G. Rehm, Mathematical modeling and computer simulation of fire phenomena, Theor. Comput. Fluid Dyn. 6 (1994) 125-139.
[67] T. Beji, J. Zhang, W. Yao, M. Delichatsios, A novel soot model for fires: validation in a laminar non-premixed flame, Combust. Flame 158 (2011) 281-290.
[68] F. Battaglia, K.B. McGrattan, R.G. Rehm, H.R. Baum, Simulating fire whirls, Combustion Theory and Modelling 4 (2000) 123.

[69] R. Zhou, Z.-N. Wu, Fire whirls due to surrounding flame sources and the influence of the rotation speed on the flame height, J. Fluid Mech. 583 (2007) 313-345.
[70] A. Yuen, G.H. Yeoh, S.C. Cheung, Q. Chan, T.B.Y. Chen, W. Yang, H. Lu, Numerical study of the development and angular speed of a small-scale fire whirl, Journal of computational science 27 (2018) 21-34.
[71] Y. Xin, J. Gore, K. McGrattan, R. Rehm, H. Baum, Large eddy simulation of buoyant turbulent pool fires, Proc. Combust. Inst. 29 (2002) 259-266.
[72] Y. Xin, J.P. Gore, K.B. McGrattan, R.G. Rehm, H.R. Baum, Fire dynamics simulation of a turbulent buoyant flame using a mixture-fraction-based combustion model, Combust. Flame 141 (2005) 329-335.
[73] Y. Xin, S. Filatyev, K. Biswas, J. Gore, R. Rehm, H. Baum, Fire dynamics simulations of a one-meter diameter methane fire, Combust. Flame 153 (2008) 499-509.
[74] K. Takagi, H. Gotoda, I.T. Tokuda, T. Miyano, Nonlinear dynamics of a buoyancy-induced turbulent fire, Physical Review E 96 (2017) 052223.
[75] K. Takagi, H. Gotoda, Effect of gravity on nonlinear dynamics of the flow velocity field in turbulent fire, Physical Review E 98 (2018) 032207.
[76] T. Yang, P. Zhang, Faster flicker of buoyant diffusion flames by weakly rotatory flows, Theor. Comput. Fluid Dyn., (2023) 1-18.
[77] T. Yang, Y. Chi, P. Zhang, Vortex interaction in triple flickering buoyant diffusion flames, Proc. Combust. Inst. 39 (2023) 1893-1903.
[78] J. Lei, N. Liu, K. Satoh, Buoyant pool fires under imposed circulations before the formation of fire whirls, Proc. Combust. Inst. 35 (2015) 2503-2510.
[79] J. Lei, N. Liu, Y. Jiao, S. Zhang, Experimental investigation on flame patterns of buoyant diffusion flame in a large range of imposed circulations, Proc. Combust. Inst. 36 (2017) 3149-3156.
[80] Y. Yang, H. Zhang, X. Xia, P. Zhang, F. Qi, An experimental study of the blue whirl onset, Proc. Combust. Inst. 39 (2023) 3705-3714.
[81] V.R. Katta, L. Goss, W.M. Roquemore, Effect of nonunity Lewis number and finite-rate chemistry on the dynamics of a hydrogen-air jet diffusion flame, Combust. Flame 96 (1994) 60-74.
[82] K. McGrattan, S. Hostikka, R. McDermott, J. Floyd, C. Weinschenk, K. Overholt, Fire dynamics simulator technical reference guide volume 1: mathematical model, NIST special publication 1018 (2013) 175.
[83] See Supplemental Material at XXX for additional information about the lumped species approach, validations of computational methods and theortical modeling, and results of flame circle arrays.
[84] J. Fang, J.-w. Wang, J.-f. Guan, Y.-m. Zhang, J.-j. Wang, Momentum-and buoyancy-driven laminar methane diffusion flame shapes and radiation characteristics at sub-atmospheric pressures, Fuel 163 (2016) 295-303.
[85] F.M. Atay, Complex time-delay systems: theory and applications, Springer2010.
[86] D.R. Reddy, A. Sen, G.L. Johnston, Time delay induced death in coupled limit cycle oscillators, Physical Review Letters 80 (1998) 5109.
[87] X. Ju, A. Bunkwang, T. Yamazaki, T. Matsuoka, Y. Nakamura, Flame Flickering can Cease Under Normal Gravity and Atmospheric Pressure in a Horizontally Moving Dual Burner System, Physical Review Applied 19 (2023) 014060.
[88] Y. Guan, K. Moon, K.T. Kim, L.K. Li, Synchronization and chimeras in a network of four ring-coupled thermoacoustic oscillators, J. Fluid Mech. 938 (2022).